\documentclass[twocolumn,letterpaper,aps,prc,superscriptaddress,nofootinbib,showpacs,floatfix]{revtex4-1}% PRC format
\usepackage{graphicx}
\usepackage{comment}
\usepackage{setspace}
\usepackage{amsmath}
\usepackage{amssymb}
\usepackage[textwidth=16cm,textheight=22cm]{geometry}
\usepackage{color}
\usepackage{indentfirst}
\usepackage{xspace}
\usepackage{hyperref}
\usepackage{verbatim}
\usepackage{epstopdf}
\usepackage{lineno}
\definecolor{red}{rgb}{1.0, 0.0, 0.0}
\definecolor{orange}{rgb}{1.0, 0.49, 0.0}
\definecolor{orangev2}{rgb}{0.91, 0.45, 0.32}   
\definecolor{blue}{rgb}{0.0, 0.0, 1.00}        
\definecolor{violetv2}{rgb}{0.57, 0.36, 0.51}

\begin{document}
%\title
\title{Modeling of Charged-Neutral Kaon Fluctuations as a Signature of DCC Production in A--A Collisions} 
\author{Ranjit~Nayak}
\email{ranjit@phy.iitb.ac.in}
\author{Sadhana~Dash}
\email{sadhana@phy.iitb.ac.in}
\author{Basanta~Kumar ~Nandi}
\email{basanta@phy.iitb.ac.in}	
\affiliation{Department of Physics, Indian Institute of Technology Bombay, Mumbai, 
 India-400076}
\author{Claude~Pruneau }
\email{pruneau@physics.wayne.edu}
\affiliation{Department of Physics and Astronomy, Wayne State University, Detroit, 48201,USA} 

\begin{abstract}

Anomalous event-by-event fluctuations of the relative yields of  neutral (K$^0_s$) and charged kaon (K$^\pm$)  have been predicted to yield a signature for the formation of Disoriented Chiral Condensate (DCC)  in relativistic heavy-ion collisions.  In this work, we model the production and decay of DCCs in the context of heavy-ion collisions at the Large Hadron Collider, and estimate the sensitivity of large acceptance detectors, such as the ALICE detector, towards the identification of such anomalous decays. Our study is based on the robust statistical observable, $\nu_{\rm dyn}$, known for its sensitivity to dynamical fluctuations. We first present simulations without DCCs, based on the HIJING and AMPT models, in order to establish
an approximate reference  for the magnitude of $\nu_{\rm dyn}({\rm K}^{\pm},{\rm K}^{0}_{s})$ and its  
centrality evolution in Pb--Pb collisions at the TeV energy scale.  We next introduce simple phenomenological 
models of K$^0_s$ vs. K$^\pm$ event-by-event yield fluctuations, which we use to study the feasibility and sensitivity of detection of the production of DCCs in heavy-ion collisions. Although the  precision
of models such as HIJING and AMPT limit their use as absolute references and thus render  anomalous fluctuations difficult to define precisely, our studies demonstrate that the  magnitude of $\nu_{\rm dyn}({\rm K}^{\pm},{\rm K}^{0}_{s})$  is in fact very sensitive to the presence of small admixture of DCCs in
normal non-DCC events. Consequently, while large values of $\nu_{\rm dyn}({\rm K}^{\pm},{\rm K}^{0}_{s})$ may not be  sufficient  to identify the existence of DCCs, nonetheless they constitute a first and necessary 
condition to signal their  possible production in  heavy-ion collisions. 
\end{abstract}
\maketitle

%%%%%%%%%%%%%%%%%%%%%%%%%%%%%%%%%%%%%%%%%%%%%%%%%%%%%%%%%%%%%%%%%%%%%
\section{Introduction}
%%%%%%%%%%%%%%%%%%%%%%%%%%%%%%%%%%%%%%%%%%%%%%%%%%%%%%%%%%%%%%%%%%%%%
Ultra-relativistic heavy-ion collisions at Large Hadron Collider (LHC)
energies produce matter consisting of  quarks and gluons  in a
de-confined state. This matter expands and cools through the critical temperature thereby forming hadrons. 
Within the hot and dense plasma
of quarks and gluons, one expects the production of  regions where the chiral symmetry 
is nearly restored. Restoration  of chiral symmetry and its
subsequent relaxation towards the normal vacuum is additionally posited to yield  transient regions 
where the value of the chiral order parameter differs from that of the surrounding
medium. These regions are called Disoriented Chiral Condensate (DCC)
\cite{dcc1,dcc2,dcc3}. Theoretical studies of the production and decay of DCCs were  formulated in the context of the SU(2) linear sigma model, more than two decades ago. It was predicted that the production and decay of DCCs shall manifest  through enhancement of electromagnetic processes and anomalous distributions of neutral and charged pion multiplicities. 
 
To date, experimental  studies of chiral symmetry restoration and searches for the production 
of DCCs were based on the two-flavor linear sigma model and involved measurements of  pion production and their fluctuations.  Unfortunately, these measurements  were largely inconclusive~\cite{wa98, minimax,star}. 
More recently, however, observation of a significant enhancement of the production of $\Omega$(and $\overline{\Omega}$)
baryons in 17 A  GeV in Pb$-$Pb collisions measured at the CERN SPS provided new insight.
It is conceivable that DCC domains also affect the strange particle
production. Strange DCC production  might then explain the observed enhancement of $\Omega$ 
observed at the SPS~\cite{omega2}. This implies that the evolution of the chiral condensate 
could feature a substantial strangeness component. Relaxation to vacuum of the DCC could
thus  produce kaon isospin fluctuations. One of the first theoretical works to address 
the observable consequence of  kaon fluctuations in the presence of many
small DCCs suggested the study of isospin fluctuations in the kaon
sector using the $\nu_{\rm dyn}$ variable~\cite{claude,kapusta}. This exploration was motivated, in part,  by 
an extension of  the two flavor
idealization of the linear sigma model to a three flavor model based on SU(3) symmetry. In this model,
 fluctuations   of  strange quarks are coupled to those of  up and
down quarks~\cite{randrup} and the chiral condensate is determined by a $u-d-s$ 
scalar field. As DCC domains relax, they  radiate pions, kaons,  and $\eta$
mesons. At low beam energy, the linear sigma model simulations indicate that pion fluctuations
should dominate three flavor DCC behaviour given the fraction of energy imparted to
kaon fluctuations is very small due to its larger mass. However, this
may not be true for collisions at LHC energies where the kaon mass is small
relative to the energy available towards particle production.  Strangeness might indeed play an important role in the decay of  DCCs  once the temperature exceeds the mass of the
strange quark. The rapid and oscillatory relaxation of order
parameters would  lead to an enhanced production and field fluctuations of
 kaons, albeit to a lesser degree than pions~\cite{kapusta}. DCC production might 
 thus induce anomalous fluctuations of the kaon total isospin measurable in the form
 of charged vs. neutral kaon yields fluctuations.  A search for charged vs. neutral kaon yield 
fluctuations at LHC energies  is thus of significant interest.

Experimental observations of DCC signals depend on various factors  such as  the probability of occurrence of DCC in a collision,  the number of DCC domains produced in an event, the 
size of the domains and the number of
particles emitted from these domains, as well as the interaction of these
particles with the rest of the collision system~\cite{wavelet}. Uncertainties about these conditions 
make the detection of DCC signals quite challenging experimentally. It is thus appropriate
to first study kaon yield fluctuations in systems that do not feature DCC production 
in order to  establish a baseline against which measurements of kaon yield fluctuations 
can be evaluated and compared.  The next step is then to introduce a simple model that
simulates, qualitatively, the behaviour of a system producing kaon DCCs. We shall vary the size and number of such DCCs relative to the full system size.

In this work, we study strangeness isospin fluctuations 
based on  measurements of the relative yield of  charged 
and neutral kaons evaluated with  the  robust correlator $\nu_{\rm dyn}(K^{\pm},K_s^0)$~\cite{claude}. 
We first  establish a baseline for the strength of this correlator with  the HIJING \cite{hijing} and
AMPT \cite{ampt}  Monte Carlo event generators. While HIJING does not produce
radial and anisotropic flows, known to be  important features of heavy-ion collisions, 
its hadron production involves a cocktail of resonances that may
give us a rough estimate of the strength of correlations between charged and neutral kaons. 
The AMPT model, on the other hand, does produce some radial and anisotropic 
flow and shall thus give us a first glimpse at the potential  impact of such   flows 
on the strength of  $\nu_{\rm dyn}(K^{\pm},K_s^0)$.  We next formulate a simple phenomenological 
model of kaon DCC production. Although somewhat simplistic, the model enables us to build
intuition as to what minimal size or maximum number of DCCs might be observed 
in heavy-ion collisions. 

Our study is designed to provide a baseline and build our intuition for a measurement 
of  $\nu_{\rm dyn}(K^{\pm},K_s^0)$  carried 
out by the ALICE collaboration with  Pb$-$Pb collisions at $\sqrt{s_{NN}}$ = 2.76 TeV~\cite{ranjitQM}. 
The definition of $\nu_{\rm dyn}(K^{\pm},K_s^0)$ in the context of this measurement and properties of this
observables are briefly discussed in Sec.~\ref{sec:Definiton}. Results of simulations carried 
out with the  HIJING and AMPT  models are presented in Sec.~\ref{sec:HijingAmptPredictions}  
whereas, the phenomenological DCC model and its predictions are discussed in Sec.~\ref{sec:DccMODELPredictions}.
This work is summarized in Sec.~\ref{sec:Summary}.

\section{STATISTICAL OBSERVABLE ($\nu_{\rm dyn}$)}
\label{sec:Definiton}

Within the context  of the SU(3) linear sigma model, DCC domains relax  by radiating pions, kaons,
 and $\eta$ mesons. The relaxation of these disoriented domains is predicted to produce widely fluctuating 
 neutral pion and kaon yields relative to those of charged pions and kaons in a given fiducial 
 acceptance. These fluctuations may be characterized in terms of 
 a neutral pion fraction $f_{\pi}$ and a neutral kaon fraction $f_{K}$.
 The neutral pion fraction is  defined as  $f_{\pi} = N_{\pi^0}/(N_{\pi^-}+N_{\pi^0}+N_{\pi^+})$
where, $N_{\pi^0}$, $N_{\pi^+}$, and $N_{\pi^-}$ represent the yields of neutral, positively charged, and negatively charged pions measured within the fiducial acceptance, respectively. The neutral kaon fraction is 
defined as  $f_{K} = (N_{K^0}+N_{\bar K^0})/(N_{K^0}+N_{\bar K^0} + N_{K^+}+N_{\bar K^-})$
where $N_{K^0}$, $N_{\bar K^0}$, $N_{K^+}$, and $N_{K^-}$ represent the yields of neutral kaon, neutral-anti-kaon, positively charged kaons, and negatively charged kaons measured within the fiducial acceptance, respectively. 
For any given DCC, the neutral pion fraction $f_{\pi}$ is predicted to fluctuate according to the probability density~\cite{dcc1}
\begin{equation}\label{eq:DccPions}
 P(f_{\pi})  = \frac{1}{2\sqrt{f_{\pi}}}, 
\end{equation}
whereas the neutral  kaon fraction $f_{K}$ shall be maximally fluctuating with a probability density~\cite {randrup}:
\begin{equation}\label{eq:DccKaons}
 P(f_{k})  = 1.  
\end{equation}
It should be noted that these probability densities deviate significantly from those
expected from  ``normal hadronic matter"  involving clusters or resonance decays, 
that  yield fluctuations of $f_{\pi}$  and $f_K$ determined by multinomial distributions  
with averages $1/3$ and $1/2$, respectively. Several  experimental difficulties arise,
however, towards measurement of these fractions. First, measurements of 
particle yields are impaired by particle losses determined by the limited
detection efficiencies of the experimental device, the reconstruction techniques, and identification
protocols used in any given analysis. This leads to binomial sampling and
a broadening of the variance of measured multiplicities. Second, the pion and kaon 
yields  may fluctuate irrespective of the production of DCCs and in response to 
fluctuations of the collision centrality. 
Third,   neutral kaons (anti-kaons) mix as weak eigenstates in the form
$K_s^0$ and $K_L^0$. The latter is difficult to observe and count on an event-by-event basis,
but the former can be easily identified and counted based on the hadronic decay mode $K_s^0\rightarrow \pi^+ \pi^-$.
 Event-by-event, half of produced neutral kaons and anti-kaons  shall yield $K_s^0$. One thus expects
 $\langle N_{K^0}\rangle = \langle N_{\bar K^0}\rangle = \langle N_{K_s^0}\rangle$. Given
 the $K_L^0$s are not measured ab-initio, the $K_s^0$ yield may thus be regarded as the total neutral kaon yield
 measured with a 50\% detection efficiency. Evidently, in practice, only those $K_s^0$ that decay into
 $\pi^+ \pi-$ will be measured and this observation will have an efficiency of its own. The number of
 neutral kaons observed on an event-by-event is thus expected to fluctuate, irrespective of the 
 production of DCCs, due to weak mixing and the finite branching ratio of the $K_s^0\rightarrow \pi^+ \pi^-$
 decay.
Fourth, and last, $K_s^0$  cannot 
be observed directly and must be reconstructed and identified statistically by means 
of topological and invariant mass cuts. Pairs of charged pions of opposite charge may be combinatorially
identified as $K_s^0$ candidates. The number of reconstructed  $K_s^0$ shall then be smeared
by the presence of this combinatorial background. 
Overall, then, it is clear that the fraction $f_K$ might fluctuate for a number of reasons having 
little to do with the existence and production of DCCs. One thus needs a fluctuation observable that is 
sensitive to the relative yields of produced charged and neutral kaons  and  robust against  losses
associated with the production (by mixing) and detection of $K_s^0$. 

Defining the number of measured charged and neutral kaons as $n_c=n_{K^+}+n_{K^-}$ and 
$n_0=n_{K_s^0}$, fluctuations of their relative values may be evaluated with the
 $\nu_{\rm dyn}(n_{c}, n_{0})$ observable defined as 
\begin{equation}
\nu_{\rm dyn}(\alpha, \beta)=R_{\alpha\alpha}+R_{\beta\beta}-2R_{\alpha\beta},
\end{equation}
where the correlators $R_{\alpha \beta}$ are calculated according to 
\begin{equation}\label{eq:rab}
R_{\alpha \beta}=\frac{\langle n_{\alpha}(n_{\alpha}-\delta_{\alpha,\beta})\rangle }{\langle n_\alpha\rangle\langle n_\beta\rangle}-1,
\end{equation}
with $\delta_{\alpha,\beta}=1$ for $\alpha = \beta$ and $\delta_{\alpha,\beta}=0$ otherwise.
It is straightforward to verify that the correlators $R_{\alpha \beta}$ are robust 
against experimental efficiencies~\cite{claude}, i.e., they are equal to the values of these correlators 
obtained based on the true (produced) multiplicities $N_c=N_{K^+}+N_{K^-}$ and 
$N_0=N_{K_s^0}$. Corrections for contamination and admixtures of combinatorial
backgrounds are possible by measuring their correlators $R_{\alpha \beta}$ explicitly.

The magnitude of  $\nu_{\rm dyn}(n_{c}, n_{0})$ is  determined by the relative strength 
of charged and neutral kaon correlations: $R_{cc}$ measures the strength of charged
kaon correlations, $R_{00}$  measures the strength of neutral  kaon correlations, and  $R_{c0}$
is sensitive to charged-neutral kaon correlations. Together, the three terms measure the strength of
fluctuations of the difference of the number of charged and neutral kaons $N_c-N_0$. Thus, $\nu_{\rm dyn}(n_{c}, n_{0})$ is thus sensitive to fluctuations of the neutral fraction $f_K$. Given it automatically accounts and corrects
physical and experimental effects causing particle losses, it then constitutes a practical observable for a measurement
of neutral vs. charged kaon yield fluctuations. 

Note that the individual terms $R_{\alpha \beta}$ shall vanish in the absence of pair correlations, i.e., for Poissonian particle production, and  their magnitude is expected  to approximately scale in inverse proportion of the total 
multiplicity of heavy-ion collisions~\cite{claude}.

\section{HIJING and AMPT MODEL PREDICTIONS }
\label{sec:HijingAmptPredictions}

We carried out  calculations of $\nu_{\rm dyn}(n_{c}, n_{0})$ for  Pb--Pb collisions  at  $\sqrt{s_{NN}}$ = 2.76 TeV  simulated with the HIJING and AMPT  event generators to establish an approximate baseline for the magnitude of charge vs. neutral kaon fluctuations to be expected in the absence of strange DCC production.  A total of $3$ Million HIJING events were generated whereas, $39$, $53$, and $39$ Million events were produced with AMPT using options (1) string-melting on (SON) and re-scattering off (ROFF), (2) string-melting off (SOFF) and re-scattering on (RON), and (3) string melting on and re-scattering on, respectively.  No momentum smearing or particle losses were used in the simulations
and $K^{0}_{S}$ were not decayed.  Events were partitioned into five centrality classes based on the fractional 
cross-section (number of events) studied as a function of  the total charged
particle multiplicity observed in the pseudo-rapidity range $2.8 \le \eta \le 5.1$ and $-3.7 \le \eta \le -1.7$.  Events were  then analyzed at generator level, in each centrality class,  by selecting  $K^{\pm}$ and $K^{0}_{S}$ in the transverse momentum range $0.2 < p_{\rm T} < 1.5$  GeV/$c$ and the pseudo-rapidity range  $|\eta| < 0.5$ to mimic the conditions of an ongoing ALICE analysis~\cite{ranjitQM}. The number of charged, $N_c$, and neutral, $N_0$, kaons were counted event-by-event and used to compute event-ensemble averages  $\langle N_c\rangle$ and $\langle N_0\rangle$, and second factorial moments 
$\langle N_c(N_c-1)\rangle$, $\langle N_0(N_0-1)\rangle$, and $\langle N_c N_0\rangle$. In turn, these were combined to compute the correlators $R_{cc}$, $R_{00}$, $R_{c0}$, and  $\nu_{\rm dyn}(N_{c}, N_{0})$ according to Eq.~(\ref{eq:rab}), in each centrality  class.
 
The HIJING and AMPT generators do not include DCC production but otherwise feature many of the physical processes required
to model heavy-ion collisions. They thus provide us a rough baseline  set of values of $\nu_{\rm dyn}$ to be expected in the 
absence of DCC formation. We  explore in the next section how the production of DCC may then modify
these basic expectations. Additionally note that given HIJING does not feature flow-like collectivity, one 
should expect its predictions of $\nu_{\rm dyn}$ to exhibit a simple $1/N$ scaling with collision centrality. It shall
then provide us with a basic reference to study the evolution of $\nu_{\rm dyn}$ with collision centrality and possible 
departure from this scaling associated with the production of DCC in mid-central to central-collisions.

\begin{figure}
\begin{center}
\includegraphics[scale=0.4]{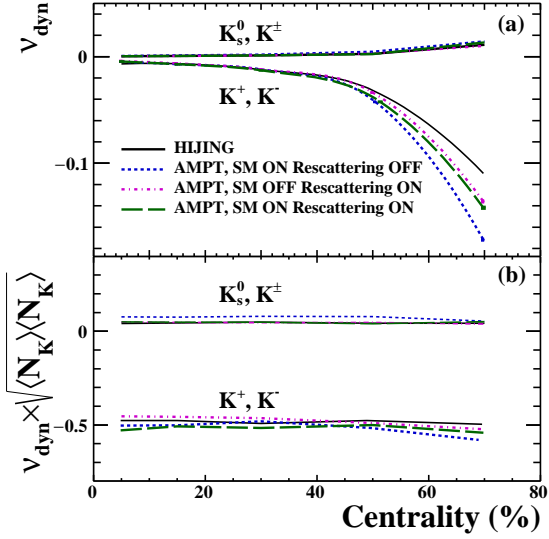}
\caption{(a) HIJING and AMPT model predictions for the collision centrality dependence of $\nu_{\rm dyn}({\rm K}^{\pm},{\rm K}^0)$ and $\nu_{\rm dyn}({\rm K}^{+},{\rm K}^{-})$ in  Pb--Pb collisions 
at $\sqrt{s_{\mathrm {NN}}}$=2.76 TeV. (b) $\nu_{\rm dyn}$ values scaled by  the geometric mean of kaon multiplicities.}
\label{fig1}
\end{center}
\end{figure}

Figure  \ref{fig1} shows a graph  of $\nu_{\rm dyn}$ vs. centrality class obtained   with
 HIJING  and  the three AMPT dynamical modes. In all four cases shown, values of $\nu_{\rm dyn}$  exhibit a 
 monotonic increase from 0-10\%, corresponding to most central collisions, to 70\%, corresponding to the
 most peripheral collisions studied in this analysis. Additionally, small differences in magnitude are observed, at any given centrality, between the three AMPT dynamical modes considered. However, their
trend with collision centrality remain similar. This indicates that
$\nu_{\rm dyn}$ is indeed sensitive to final state interactions that modify the  yield
of produced $K^{\pm}$ and $K^{0}_{S}$ as well as their correlations in Pb--Pb collisions, even if these effects are small.

In order to account for the dilution of the $R_2$ correlators with increasing particle multiplicity (and number of sources), we 
scale values of  $\nu_{\rm dyn}$, in Fig.~\ref{fig1}~(b),  by the geometric mean, 
$\sqrt{N_{{\rm K}^{\pm}}N_{{\rm K}^0}}$, of the numbers of $K^{\pm}$ and $K^{0}_{S}$ 
produced in the fiducial acceptance of the measurement.   As expected, scaled values calculated with HIJING and AMPT/RON are essentially invariant with collision centrality, whereas  scaled values calculated with  AMPT/SON/ROFF exhibit a weak collision centrality dependence, owing to the production of higher mass states in most central collisions.  Overall, both HIJING and AMPT produce nearly  invariant 
values of $\nu_{\rm dyn}\sqrt{N_{{\rm K}^{\pm}}N_{{\rm K}^0}}$  showing that the expected $1/N$ dilution is approximately verified.

It is interesting to compare calculations of $\nu_{\rm dyn}(N_{c}, N_{0})$, discussed above, with 
predictions for $\nu_{\rm dyn}({N}_{+},{N}_{-})$ computed with the same models and also shown in Fig.~\ref{fig1}~(a). Conservation laws, namely electric charge  and strangeness conservation, restrict the level of fluctuations possible in the production of charged kaons. The cumulant  $R_{+-}$ is  larger than either of $R_{++}$ or $R_{--}$)  and $\nu_{\rm dyn}(N_{+},N_{-})$ is consequently negative at all collision centralities. By contrast, charge conservation does not limit the fluctuations of the relative yields of K$^{\pm}$ and  K$^0$ and $\nu_{\rm dyn}(N_{c},N_{0})$ takes positive values at all centralities. Note, however, that the magnitude of $\nu_{\rm dyn}(N_{+},N_{-})$ influences in part the 
value of $\nu_{\rm dyn}(N_{c},N_{0})$, given
\begin{align} \nonumber
R_{cc} &=f^{2}R_{++} + (1-f)^{2}R_{--} + 2f(1-f)R_{+-} \\ 
&= \frac{1}{4}\left[ 2R_{++} + 2R_{--} - \nu_{\rm dyn}(N_{+},N_{-}) \right], 
\end{align}
where $f=\langle N_+\rangle/(\langle N_+\rangle + \langle N_-\rangle)$ and in the second line, we assumed $f=0.5$, which is approximately valid at LHC energy.
Finally, note that values of $\nu_{\rm dyn}({\rm K}^{+},{\rm K}^{-})$ scaled by the number of charged kaons are also invariant with collision centrality, while AMPT/SON/ROFF displays 
a modest collision centrality dependence. 

\section{DCC  MODEL SIMULATIONS}
\label{sec:DccMODELPredictions}

We simulate the production of DCCs with a simple phenomenological model to gain insight into the correlation strength measurable with $\nu_{\rm dyn}$ in the presence of condensates decaying into neutral and charged kaons. The DCC simulator is designed to match the pion and kaon multiplicity  production observed experimentally and  involves ``normal" and DCC particle 
generation components. The former produces charged and neutral particles based on a binomial  distribution, whereas the DCC particle production is determined by Eqs.~(\ref{eq:DccPions},~\ref{eq:DccKaons}) for pions and kaons, respectively. 
Events can be set to consist of DCC matter entirely or a mix of normal and DCC matter by used of a parameter $f_{\rm DCC}$ which controls the number of particles produced within the binomial and DCC components. The average  fraction of particles consisting of kaons  is determined by a user selected parameter $f_K$ set at simulation startup. However, the kaon to pion yield ratio is allowed
to fluctuate according to a binomial distribution determined by this fraction and the total multiplicity.  This multiplicity is randomly chosen according to a PDF that approximately replicates the charged particle multiplicity reported by the ALICE collaboration in 0-10\%
collision centrality,  scaled by a factor of 3/2 to account for neutral particle production.  The number of $K^{0}_{S}$s in the DCC and 
normal parts of an event are  randomly generated
according to a binomial distribution with  a mean probability of 0.5 based on the number of neutral kaons in either parts.

\begin{figure}
\begin{center}
\includegraphics[scale=0.4]{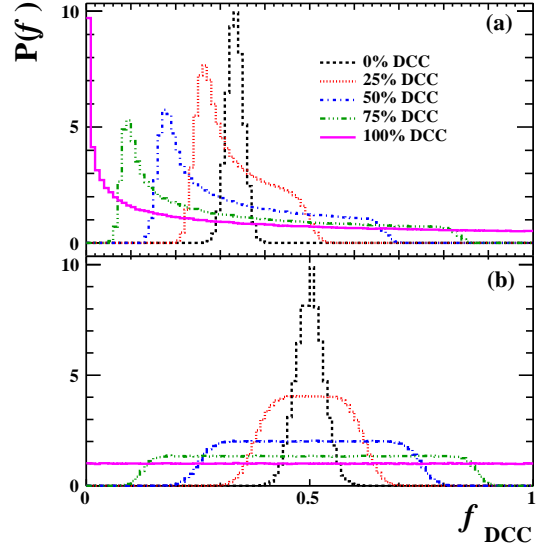}
\caption{Probability densities of the fractions of (a) neutral pions  and (b) neutral kaons  
for selected values of the fraction $f_{\rm DCC}$ of particles produced within DCCs. }
\label{fig2}
\end{center}
\end{figure}

\begin{figure}
\begin{center}
\includegraphics[scale=0.4]{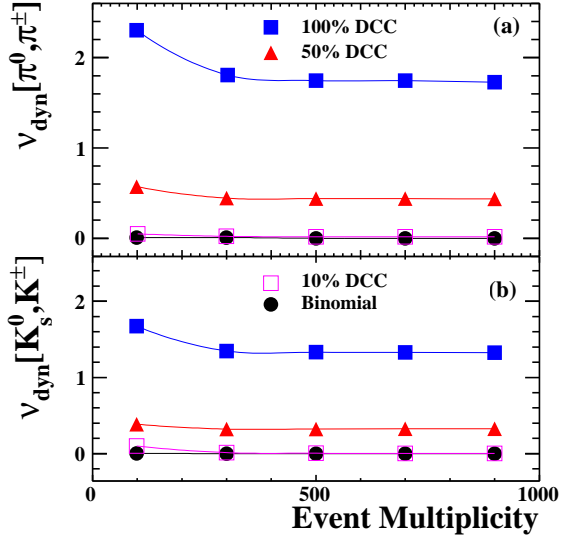}
\caption{Evolution of (a) $\nu_{\rm dyn}(\pi^0,\pi^{\pm})$ and (b)  $\nu_{\rm dyn}({\rm K}^0,{\rm K}^{\pm})$, with event multiplicity for  selected values of the fraction of particles produced within DCCs.}
\label{fig3}
\end{center}
\end{figure}

\begin{figure}
\begin{center}
\includegraphics[scale=0.4]{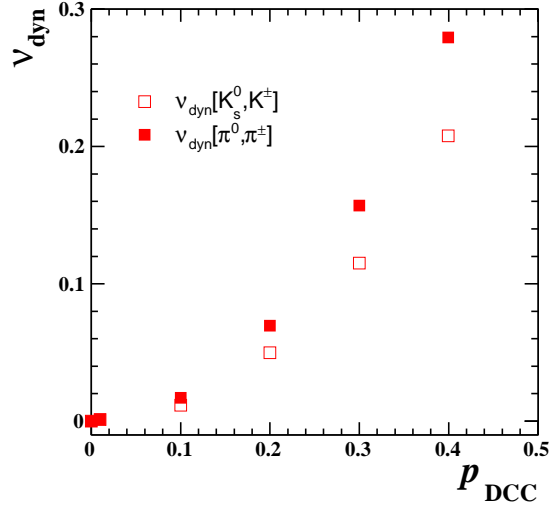}
\caption{$\nu_{\rm dyn}$ as a function of percentage of DCC-like events ($p_{\rm DCC}$) for a given multiplicity bin.}
\label{fig4}
\end{center}
\end{figure}

\begin{figure}
\begin{center}
\includegraphics[scale=0.4]{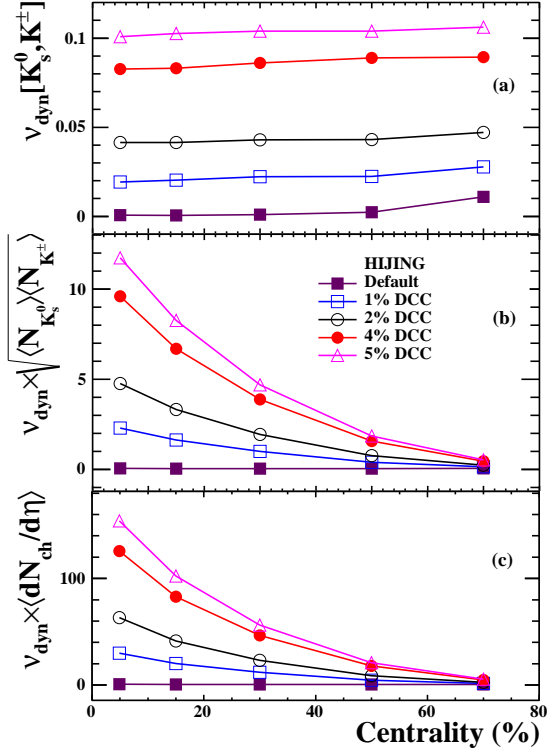}
\caption{(a): $\nu_{\rm dyn}$ for $K_{s}^{0}-K^{\pm}$ as a function of  centrality, (b): $\nu_{\rm dyn}$ scaled with kaon multiplicity as a function of centrality, and (c): $\nu_{\rm dyn}$ scaled with charged particle multiplicity as a function of centrality, for Pb$-$Pb collisions 
at $\sqrt{s_{\mathrm {NN}}}$ = 2.76 TeV.}
\label{figflipping}
\end{center}
\end{figure}

We explore the impact of DCC production by varying the size and number of DCC in generated events. Figure~\ref{fig2} displays the  fractions of neutral pions (top panel) and  kaons (bottom panel) obtained in generated events with  selected values of the DCC fraction $f_{\rm DCC}$. 
%Figure~\ref{fig3} presents plots of the evolution of $\nu_{\rm dyn}(\pi^{0},\pi^{\pm})$ (top) and $\nu_{\rm dyn}({\rm K}^{0},{\rm K}^{\pm})$ (bottom) with the event multiplicity . In the context of the simple model used to generate the shown distributions, the DCC fraction is fixed, i.e., independent of the event multiplicity. 
 We consider few distinct  particle production scenarios: 
{\bf Scenario 1}: All generated events  are assumed to contain one DCC domain but the size of the  domains is varied by changing the fraction of (a) pions  and (b) kaons they produce relative to the
full system. As shown in Fig.~\ref{fig3},   $\nu_{\rm dyn}$  increases monotonically with the DCC size and reaches  very large magnitudes for  events containing large DCC domains. The rise of  $\nu_{\rm dyn}$  at small event multiplicity is due to a  decline of the cross-term $R_{c0}$ relative to $R_{00}$ and $R_{cc}$.
{\bf Scenario 2} : The DCC size is kept fixed and accounts for 100 percent of the event size but the probability of occurrence of DCCs is  varied. The magnitude of  $\nu_{\rm dyn}$ rises with the fraction of events containing a DCC, called $p_{\rm DCC}$, as illustrated in Fig.~\ref{fig4}. Sizeable $\nu_{\rm dyn}$ magnitudes and deviations from  binomial expectation values occur for probabilities as small as $p_{\rm DCC}=0.01$. 
{\bf Scenario 3} : We use HIJING events as a baseline for the description of pion and kaon production and their event-by-event fluctuations. DCC-like fluctuations are introduced by randomizing the charge of 
kaons produced by HIJING. Figure~\ref{figflipping} displays the collision centrality evolution of $\nu_{\rm dyn}$
obtained when progressively increasing the fraction of events containing a single (large) strange DCC. As already discussed in Sec. ~\ref{sec:HijingAmptPredictions}, the magnitude $\nu_{\rm dyn}$ exhibits an approximate $1/N$ behaviour with increasing event multiplicity $N$, shown as a purple line in Fig.~\ref{figflipping}. Injection of
DCC-like fluctuations, however, drastically changes this behaviour and the distribution of $\nu_{\rm dyn}$
with collision centrality becomes progressively flat (nearly invariant) when rising the fraction of events containing DCC-like fluctuations. Values of $\nu_{\rm dyn}$ scaled by the geometric mean of the number of kaons, shown in Fig.~\ref{figflipping} (b), and the charged particle density, shown in Fig.~\ref{figflipping} (c), are no longer invariant with collision centrality and sharply  rise in central Pb--Pb collisions. Note that significant deviations from the HIJING baseline are found already when the fraction of DCC amounts to only one percent. The magnitude and collision centrality evolution of $\nu_{\rm dyn}$ are thus indeed
quite sensitive to the presence of DCC-like fluctuations in Pb--Pb collisions.

\section{Summary}
\label{sec:Summary}

The event-by-event fluctuations of neutral and charged kaons were studied using the 
$\nu_{\rm dyn}$ observable in Monte Carlo models (HIJING and AMPT) in
Pb$-$Pb collisions at $\sqrt{s_{NN}}$ = 2.76 TeV for various collision  centralities.  The
estimates from these models are relevant to the ongoing experimental
search of DCC-like signals in heavy ion collisions at LHC and RHIC as
these models do not include the dynamics of DCC physics. 
A simple phenomenological model was developed to implement the DCC-type events
and study the sensitivity of the $\nu_{\rm dyn}$ observable to the
fraction of DCC events as well as the size of DCC domains. The value of $\nu_{\rm dyn}$ increases with a rising fraction of DCC-like events in the sample for a given multiplicity class.
The variation of $\nu_{\rm dyn}$ was also studied as a function of multiplicity for different size of DCC domains. Both  studies indicate that
$\nu_{\rm dyn}$ is very sensitive to the 
presence of DCC-like events in heavy-ion collisions.  
 
\section{Acknowledgements}
This work  was supported in part by the Department of Science and Technology (DST), Government of India and the United States Department of Energy, Office of Nuclear Physics (DOE NP), United States of America, under grant No.  DE-FG02-92ER40713.

\end{document}